\documentclass[journal]{IEEEtran}
\usepackage[utf8]{inputenc}
\usepackage{color}
\usepackage{array}
\usepackage{verbatim}
\usepackage{float}
\usepackage{amsmath}
\usepackage{amsthm}
\usepackage{amssymb}
\usepackage{graphicx}
\usepackage{longtable}
\usepackage{multirow}
\usepackage[unicode=true,
bookmarks=false,
breaklinks=false,pdfborder={0 0 1},colorlinks=false]
{hyperref}
\hypersetup{
	colorlinks,bookmarksopen,bookmarksnumbered,citecolor=blue,urlcolor=blue}
\usepackage{cite}

\floatstyle{ruled}
\newfloat{algorithm}{tbp}{loa}
\providecommand{\algorithmname}{Algorithm}
\floatname{algorithm}{\protect\algorithmname}

\makeatletter
\let\oldforeign@language\foreign@language
\DeclareRobustCommand{\foreign@language}[1]{%
	\lowercase{\oldforeign@language{#1}}}

\let\oldforeign@language\foreign@language
\DeclareRobustCommand{\foreign@language}[1]{%
	\lowercase{\oldforeign@language{#1}}}

\ifCLASSINFOpdf
\else
\fi

\makeatletter

\newtheorem{rem}{Remark}

\newtheorem{assum}{Assumption}

\begin{document}
	\bstctlcite{IEEEexample:BSTcontrol}

	\title{FLC tuned with Gravitational Search Algorithm for Nonlinear Pose Filter}
	
	\author{Trenton S. Sieb, Ajay Singh, Lorelei Guidos, and~Hashim~A.~Hashim\\% <-this % stops a space
		Software Engineering\\
		Department of Engineering and Applied Science\\
		Thompson Rivers University,	Kamloops, British Columbia, Canada, V2C-0C8\\
		siebt19@mytru.ca, ludhera17@mytru.ca, guidosl16@mytru.ca, and hhashim@tru.ca
		% \thanks{
		%	Name 1, Name 2, Name 3 and H. A. Hashim  are with Software Engineering, Department of Engineering and Applied Science, Thompson Rivers University, Kamloops, British Columbia, Canada, V2C-0C8, e-mail: email1@mytru.ca, email2@mytru.ca, email3@mytru.ca, and hhashim@tru.ca.
		%\\This research work was supported by Thompson Rivers University Internal research fund # 102034.
		% }
		\thanks{This work was supported in part by Thompson Rivers University Internal
			research fund, RGS-2020/21 IRF, \# 102315.}
		
	}
	%\author{Hashim~A.~Hashim$^*$,~\IEEEmembership{~Member, IEEE}, Lyndon J. Brown, and~Kenneth McIsaac,~\IEEEmembership{~Fellow, IEEE}% <-this % stops a space
	%\thanks{$^*$Corresponding author, H. A. Hashim, L. J. Brown and K. McIsaac are with the Department of Electrical and Computer Engineering,
	%University of Western Ontario, London, ON, Canada, N6A-5B9, e-mail: hmoham33@uwo.ca, lbrown@uwo.ca and kmcisaac@uwo.ca.}}
	
	%\markboth{--,~Vol.~-, No.~-, \today}{Hashim \MakeLowercase{\textit{et al.}}: Nonlinear Pose Filters on SE(3) with Guaranteed Transient and Steady-state Performance}
	%\markboth{}{Hashim \MakeLowercase{\textit{et al.}}: Guaranteed Performance Nonlinear Estimator for SLAM}
	
	\maketitle
	
	\begin{abstract}
		Nonlinear pose (\textit{i.e,} attitude and position) filters are characterized
		with simpler structure and better tracking performance in comparison
		with other methods of pose estimation. A critical factor when designing
		a nonlinear pose filter is the selection of the error function. Conventional
		design of nonlinear pose filter design trade-off between fast adaptation
		and robustness. This paper introduces a new practical approach based
		on fuzzy rules for on-line continuous tuning of the nonlinear pose
		filter. Each of input and output membership functions are optimally
		tuned using graphical search algorithm optimization considering both
		pose error and its rate of change. The proposed approach is characterized
		with high adaptation features and strong level of robustness. Therefore,
		the proposed approach results of robust and fast convergence properties.
		The simulation results show the effectiveness of the proposed approach
		considering uncertain measurements and large error in initialization.
	\end{abstract}
	
	%\begin{IEEEkeywords}
	%Nonlinear filter algorithm, Simultaneous Localization and Mapping, asymptotic stability, systematic convergence,
	%pose, attitude, position, landmark, adaptive estimate, SLAM, SE(3),
	%SO(3).
	%\end{IEEEkeywords}

	\IEEEpeerreviewmaketitle{}

	\section{Introduction}
	
	\IEEEPARstart{F}{rom} space craft and satellites to vehicles underwater and in the
	air, knowing an accurate three-dimensional estimate of the pose of
	a rigid-body is a task that is vitally important \cite{Hashim2020SLAMIEEELetter,markley2003attitude,hashim2018SO3Stochastic,hashim2020SO3Wiley}.
	The pose consists of two components: 1) orientation and 2) position.
	Attitude is often used in place of orientation so they will be used
	interchangeably. Attitude reconstruction is often done using an algebraic
	approach. This involves using an algorithm such as QUEST \cite{shuster1981three}
	or singular value decomposition (SVD) \cite{markley1988attitude}
	that takes two or more noncollinear inertial frame vectors as well
	as the object’s body frame vectors. However, this method is quite
	susceptible to noise and bias which can damage results and make them
	unusable. This effect is particularly potent if the rigid-body is
	equipped with a low cost IMU.
	
	An alternative method that has been used historically to address the
	issue of estimating the attitude is a Gaussian or nonlinear deterministic
	filter. The Kalman filter (KF), extended KF (EKF), and multiplicative
	EKF (MEKF) are all examples within the family of Gaussian filters
	that consider the unit quaternion to represent attitude \cite{hashim2018SO3Stochastic,hashim2020SO3Wiley}.
	However, the attitude problem being of a nonlinear nature, gives nonlinear
	deterministic attitude filters evolved on the Special Orthogonal Group
	$SO\left(3\right)$, an acute advantage over Gaussian filters. They
	outperform in many aspects such as with simpler derivation and representation,
	reduced processing power, and improved tracking convergence \cite{hashim2018SO3Stochastic,mahony2008nonlinear,mohamed2019filters}. 
	
	In order to ensure accurate position estimation, a good attitude estimation
	is required. Furthermore, as with attitude, it is better to use a
	nonlinear approach, using filters evolved on the Special Euclidean
	Group $SE\left(3\right)$ \cite{hashim2020SE3Stochastic,hashim2019SE3Det,baldwin2007complementary,hashim2018SE3Stochastic}.
	These filters require measurements that can be derived from a group
	velocity vector, on-board vectorial measurements from a device such
	as an IMU, landmark measurements from a builtin vision system, and
	an estimate of the bias in the measured velocity \cite{hashim2019SE3Det,hashim2018SE3Stochastic,hashim2020SE3Stochastic}.
	Nonlinear filters are commonly used along with a computer vision system
	with a monocular camera and an IMU. In \cite{baldwin2007complementary},
	the pose filter was developed on $SE\left(3\right)$ with a proven
	exponentially stable performance. Although the implementation of said
	filter requires pose reconstruction, modifications can be made that
	allow function using purely a set of vectorial measurements \cite{baldwin2009nonlinear,hua2011observer},
	which adds simplicity and avoids the reconstruction. However, despite
	the simplicity in \cite{baldwin2007complementary,baldwin2009nonlinear,hua2011observer},
	the results of data collected show that these filters are highly sensitive
	to noise in the measurements. Additionally, the conventional design
	of pose filters \cite{baldwin2009nonlinear,hua2011observer} is characterized
	with slow convergence of tracking error.
	
	Fuzzy logic controller (FLC) is an intelligent approach which showed
	essential solutions in several control applications, for example,
	$\mathcal{L}_{1}$ adaptive controller tuned with FLC \cite{hashim2015L1},
	adaptive fuzzy controller for mobile robots \cite{shi2019fuzzy},
	and others. Also, evolutionary techniques went through accelerated
	developments over the last few decades. They have the capability to
	be an optimal fit with a wide range of control applications such as
	the gravitational search algorithm (GSA) which was proposed as a global
	search technique in \cite{rashedi2009gsa}. The necessity to tune
	originally fixed coefficients of controllers and filters has been
	widely used in various applications, such as \cite{hashim2015L1,yu2017fuzzy}.
	Moreover, they have fundamental role in data mining \cite{eltoukhy2019data,eltoukhy2018joint,eltoukhy2019robust}.
	
	Thereby, this study presents fuzzy tuning the gain of the nonlinear
	pose filter where the fuzzy input and output membership functions
	are optimized using GSA, considering the pose error and its rate of
	change. The FLC-based tuning is an on-line carried out during the
	estimation process. GSA determines the optimal values of input and
	output membership functions through off-line tuning. FLC is introduced
	to enhance the trade-off between robustness and fast convergence.
	The gain of the nonlinear pose filter is dynamically tuned, hence
	resulting in better performance. Actually, the proposed technique
	solves the dilemma of fast adaptation and fast convergence response.
	The proposed method is simpler and can be easily implemented.
	
	The rest of the paper is organized as follows: Section \ref{sec:SE3PPF_Math-Notations}
	presents a short overview of numerical and mathematical representations
	of $SO\left(3\right)$ and $SE\left(3\right)$ parameterization. Section
	\ref{sec:Pose_Problem-Formulation-in} articulates the pose problem,
	demonstrates the filter structure and error criteria, and presents
	the nonlinear structure of the attitude filter. Section \ref{sec:Filter_Strategy}
	presents the proposed filter strategy, which includes a brief introduction
	of the gravitational search algorithm, fuzzy logic controller and
	a diagram of the implementation process. Section \ref{sec:Results}
	shows the obtained results and validates the robustness of the proposed
	filters. Finally, Section \ref{sec:SO3PPF_Conclusion} completes the
	work with concluding comments.
	
	\section{{\normalsize{}Preliminaries of $\mathbb{SE}\left(3\right)$ \label{sec:SE3PPF_Math-Notations}}}
	
	In this paper $\left\{ \mathcal{B}\right\} $ denotes body-frame of
	a reference and $\left\{ \mathcal{I}\right\} $ denotes the inertial-frame
	of a reference. $\left\Vert x\right\Vert =\sqrt{x^{\top}x}$ denotes
	Euclidean norm for all $x\in\mathbb{R}^{p}$. $\mathbf{I}_{p}$ denotes
	a $p$-by-$p$ identity matrix. $R\in\left\{ \mathcal{B}\right\} $
	denotes an orientation of a rigid-body in the space which is commonly
	termed attitude. Define $SO\left(3\right)$ as the Special Orthogonal
	Group
	\[
	SO\left(3\right)=\left\{ \left.R\in\mathbb{R}^{3\times3}\right|RR^{\top}=R^{\top}R=\mathbf{I}_{3}\text{, }{\rm det}\left(R\right)=+1\right\} 
	\]
	with $\mathbf{I}_{3}$ being a $3$-by-$3$ identity matrix and ${\rm det\left(\cdot\right)}$
	is a determinant. Let $SE\left(3\right)$ be the Special Euclidean
	Group where
	\[
	SE\left(3\right)=\left\{ \left.\boldsymbol{H}\in\mathbb{R}^{4\times4}\right|R\in SO\left(3\right),P\in\mathbb{R}^{3}\right\} 
	\]
	Also, $\boldsymbol{H}\in SE\left(3\right)$ is commonly known as the
	homogeneous transformation matrix that expresses the pose of the rigid-body
	as below
	\begin{equation}
	\boldsymbol{H}=\left[\begin{array}{cc}
	R & P\\
	0_{1\times3} & 1
	\end{array}\right]\in SE\left(3\right)\label{eq:Pose_T_matrix}
	\end{equation}
	where $P$ and $R$ are position and attitude, respectively, and $0_{1\times3}$
	being a zero row. The Lie-algebra related to $SO\left(3\right)$ is
	defined as $\mathfrak{so}\left(3\right)$ and is given by
	\[
	\mathfrak{so}\left(3\right)=\left\{ \left.X\in\mathbb{R}^{3\times3}\right|X^{\top}=-X\right\} 
	\]
	where $X$ is a skew symmetric matrix. Consider the map $\left[\cdot\right]_{\times}:\mathbb{R}^{3}\rightarrow\mathfrak{so}\left(3\right)$
	to be
	\[
	\left[x\right]_{\times}=\left[\begin{array}{ccc}
	0 & -x_{3} & x_{2}\\
	x_{3} & 0 & -x_{1}\\
	-x_{2} & x_{1} & 0
	\end{array}\right]\in\mathfrak{so}\left(3\right),\hspace{1em}x=\left[\begin{array}{c}
	x_{1}\\
	x_{2}\\
	x_{3}
	\end{array}\right]
	\]
	For any $\mathcal{Y}=\left[y_{1}^{\top},y_{2}^{\top}\right]^{\top}$
	with $y_{1},y_{2}\in\mathbb{R}^{3}$, one has
	\[
	\left[\mathcal{Y}\right]_{\wedge}=\left[\begin{array}{cc}
	\left[y_{1}\right]_{\times} & y_{2}\\
	0_{1\times3} & 0
	\end{array}\right]\in\mathfrak{se}\left(3\right)
	\]
	$\mathfrak{se}\left(3\right)$ is the Lie algebra of $SE\left(3\right)$
	given by{\small{}
		\begin{align*}
		\mathfrak{se}\left(3\right) & =\left\{ \left.\left[\mathcal{Y}\right]_{\wedge}\in\mathbb{R}^{4\times4}\right|\exists y_{1},y_{2}\in\mathbb{R}^{3}:\left[\mathcal{Y}\right]_{\wedge}=\left[\begin{array}{cc}
		\left[y_{1}\right]_{\times} & y_{2}\\
		0_{1\times3} & 0
		\end{array}\right]\right\} 
		\end{align*}
		Also, for $x,y\in\mathbb{R}^{3}$ and $x_{0},y_{0}\in\mathbb{R}$
		consider the following definition
		\[
		\left[\begin{array}{c}
		x\\
		x_{0}
		\end{array}\right]\wedge\left[\begin{array}{c}
		y\\
		y_{0}
		\end{array}\right]=\left[\begin{array}{c}
		x\times y\\
		x_{0}y-y_{0}x
		\end{array}\right]\in\mathbb{R}^{6}
		\]
		For more details on $SO\left(3\right)$ visit \cite{hashim2018SO3Stochastic,hashim2020SO3Wiley}
		and $SE\left(3\right)$ visit \cite{hashim2019SE3Det,hashim2020SE3Stochastic}.}{\small\par}
	
	\section{{\normalsize{}Problem Formulation \label{sec:Pose_Problem-Formulation-in}}}
	
	This section aims to present the pose problem, pose measurements,
	error criteria and nonlinear pose filter design.
	
	\subsection{{\normalsize{}Pose Dynamics and Measurements\label{subsec:SE3PPF_Pose-Kinematics}}}
	
\begin{figure}[h]
		\centering{}\includegraphics[scale=0.3]{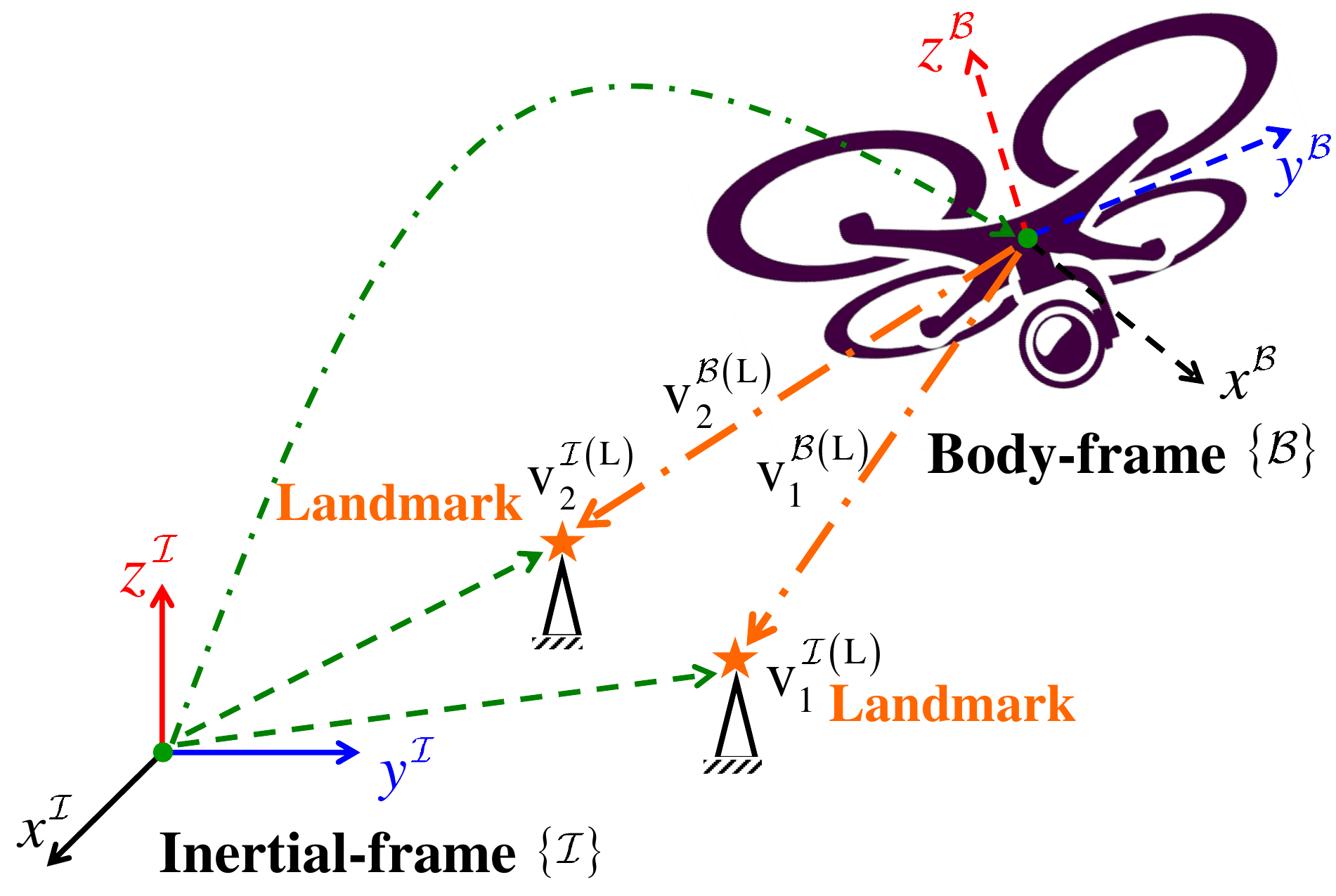}\caption{Pose filtering problem of a rigid-body \cite{hashim2019Conf1,hashim2019SE3Det}.}
		\label{fig:SE3PPF_1} 
	\end{figure}

	As mentioned earlier, pose of a rigid-body in the space consists of
	two components: orientation (attitude) and position. Attitude is given
	by $R\in\mathbb{SO}\left(3\right)$ while position is defined as $P\in\mathbb{R}^{3}$.
	Note that $R\in\left\{ \mathcal{B}\right\} $ and $P\in\left\{ \mathcal{I}\right\} $.
	The pose filtering problem is depicted in Fig. \ref{fig:SE3PPF_1}
	
	Let the superscripts $\mathcal{B}$ and $\mathcal{I}$ be elements
	related to $\left\{ \mathcal{B}\right\} $ and $\left\{ \mathcal{I}\right\} $,
	respectively. Attitude can be represented, given $n$ observations
	in the inertial-frame and their measurements in the body-frame, where
	the $i$th body-frame vector measurement is defined by
	\begin{equation}
	{\rm v}_{i}^{\mathcal{B}}=R^{\top}{\rm v}_{i}^{\mathcal{I}}+b_{i}^{\mathcal{B}}+n_{i}^{\mathcal{B}}\in\mathbb{R}^{3}\label{eq:SE3STCH_Vect_R}
	\end{equation}
	where ${\rm v}_{i}^{\mathcal{I}}$ is a known observation, $b_{i}^{\mathcal{B}}$
	is unknown bias, and $n_{i}^{\mathcal{B}}$ is unknown noise for all
	$i=1,2,\ldots,n$. The equivalent normalization of ${\rm v}_{i}^{\mathcal{I}}$
	and ${\rm v}_{i}^{\mathcal{B}}$ in Eq. \eqref{eq:SE3STCH_Vect_R}
	is expressed as
	\begin{equation}
	\upsilon_{i}^{\mathcal{I}}=\frac{{\rm v}_{i}^{\mathcal{I}}}{\left\Vert {\rm v}_{i}^{\mathcal{I}}\right\Vert },\hspace{1em}\upsilon_{i}^{\mathcal{B}}=\frac{{\rm v}_{i}^{\mathcal{B}}}{\left\Vert {\rm v}_{i}^{\mathcal{B}}\right\Vert }\label{eq:SE3STCH_Vector_norm}
	\end{equation}
	The position of a moving vehicle can be reestablished for a known
	$R$ and $L$ known landmarks. The $i$th landmark measurement in
	the body-frame can be expressed as \cite{hashim2019SE3Det,hashim2020SE3Stochastic}
	\begin{equation}
	y_{i}^{\mathcal{B}}=R^{\top}\left(p_{i}^{\mathcal{I}}-P\right)+\bar{b}_{i}^{\mathcal{B}}+\bar{n}_{i}^{\mathcal{B}}\in\mathbb{R}^{3}\label{eq:SE3STCH_Vec_Landmark}
	\end{equation}
	with $p_{i}^{\mathcal{I}}$ being a known landmark, $\bar{b}_{i}^{\mathcal{B}}$
	being unknown bias, and $\bar{n}_{i}^{\mathcal{B}}$ being unknown
	noise for all $i=1,2,\ldots,L$. 
	\begin{assum}
		\label{Assum:SE3STCH_1} (Pose observability) The pose of a rigid
		body is observable if one of the following three cases is met
		
		\textbf{case 1}. At least two non-collinear vectors in Eq. \eqref{eq:SE3STCH_Vector_norm}
		and one landmark in Eq. \eqref{eq:SE3STCH_Vec_Landmark} are available.
		
		\textbf{case 2}. At least one non-collinear vectors in Eq. \eqref{eq:SE3STCH_Vector_norm}
		and two landmark in Eq. \eqref{eq:SE3STCH_Vec_Landmark} are available.
		
		\textbf{case 3}. At least three landmarks in Eq. \eqref{eq:SE3STCH_Vec_Landmark}
		are available.
	\end{assum}
	The dynamics of the homogeneous transformation matrix in Eq. \eqref{eq:Pose_T_matrix}
	is given by
	\begin{align}
	\dot{\boldsymbol{H}} & =\boldsymbol{H}\left[\mathcal{Y}\right]_{\wedge}\label{eq:Pose_T_Dynamics}
	\end{align}
	such that $\dot{R}=R\left[\Omega\right]_{\times}$, $\dot{P}=RV$,
	$\mathcal{Y}=\left[\Omega^{\top},V^{\top}\right]^{\top}\in\mathbb{R}^{6}$
	denotes a group velocity vector. Also, $\Omega\in\mathbb{R}^{3}$
	is the true angular and $V\in\mathbb{R}^{3}$ is the translational
	velocity. Unfortunately, the measured velocity vector is corrupted
	by uncertain components
	\begin{align}
	\mathcal{Y}_{m} & =\mathcal{Y}+b+\omega\in\left\{ \mathcal{B}\right\} \label{eq:SE3PPF_Angular}
	\end{align}
	where $\mathcal{Y}_{m}=\left[\Omega_{m}^{\top},V_{m}^{\top}\right]^{\top}$,
	$b=\left[b_{\Omega}^{\top},b_{V}^{\top}\right]^{\top}$, and $\omega=\left[\omega_{\Omega}^{\top},\omega_{V}^{\top}\right]^{\top}$
	such that $b_{\Omega},b_{V}\in\mathbb{R}^{3}$ are the unknown constant
	bias and $\omega_{\Omega},\omega_{V}\in\mathbb{R}^{3}$ are unknown
	noise. Let the estimate of $\boldsymbol{H}$ in Eq. \eqref{eq:Pose_T_matrix}
	be
	\begin{equation}
	\hat{\boldsymbol{H}}=\left[\begin{array}{cc}
	\hat{R} & \hat{P}\\
	0_{1\times3} & 1
	\end{array}\right]\label{eq:SE3PPF_Test_matrix}
	\end{equation}
	where $\hat{R}$ and $\hat{P}$ are the estimates of $R$ and $P$,
	respectively. Let the error between $\boldsymbol{H}$ and $\hat{\boldsymbol{H}}$
	be
	\begin{align}
	\tilde{\boldsymbol{H}} & =\hat{\boldsymbol{H}}\boldsymbol{H}^{-1}=\left[\begin{array}{cc}
	\tilde{R} & \tilde{P}\\
	0_{1\times3} & 1
	\end{array}\right]\label{eq:SE3PPF_Terr_matrix}
	\end{align}
	
	\noindent with $\tilde{R}=\hat{R}R^{\top}$ and $\tilde{P}=\hat{P}-\tilde{R}P$
	being errors in attitude and position, respectively. This work aims
	to drive $\hat{\boldsymbol{H}}\rightarrow\boldsymbol{H}$ with fast
	adaptation and high measures of robustness to ensure that $\tilde{P}\rightarrow0_{3\times1}$,
	$\tilde{R}\rightarrow\mathbf{I}_{3}$, and $\tilde{\boldsymbol{H}}\rightarrow\mathbf{I}_{4}$.
	
	\subsection{Nonlinear Pose Filter Design \label{subsec:Filter_Design}}
	
	The filter design in this Section follows the structure in \cite{hua2011observer}
	where the contribution is the introduction of adaptively tuned gain.
	Consider the following filter design \cite{hua2011observer} 
	\begin{align}
	& \left[\begin{array}{c}
	x\\
	x_{0}
	\end{array}\right]\wedge\left[\begin{array}{c}
	y\\
	y_{0}
	\end{array}\right]=\left[\begin{array}{c}
	x\times y\\
	x_{0}y-y_{0}x
	\end{array}\right]\in\mathbb{R}^{6},\,x,y\in\mathbb{R}^{3}\nonumber \\
	\mathbf{U}= & \frac{1}{2}\sum_{i=1}^{N_{{\rm L}}}s_{i}^{{\rm L}}\hat{\boldsymbol{T}}\left[\begin{array}{c}
	y_{i}^{\mathcal{B}}\\
	1
	\end{array}\right]\wedge\left[\begin{array}{c}
	p_{i}^{\mathcal{I}}\\
	1
	\end{array}\right]\nonumber \\
	& +\frac{1}{2}\sum_{i=1}^{N_{{\rm R}}}s_{i}^{{\rm R}}\hat{\boldsymbol{T}}\left[\begin{array}{c}
	{\rm v}_{i}^{\mathcal{B}}\\
	0
	\end{array}\right]\wedge\left[\begin{array}{c}
	{\rm v}_{i}^{\mathcal{I}}\\
	0
	\end{array}\right]\nonumber \\
	\dot{\hat{\boldsymbol{T}}}= & \hat{\boldsymbol{T}}\left[\mathcal{Y}_{m}-\hat{b}+\boldsymbol{K}W\right]_{\wedge}\nonumber \\
	\dot{\hat{b}}= & -\gamma\left[\begin{array}{cc}
	\hat{R}^{\top} & 0_{3\times3}\\
	-\hat{R}^{\top}\left[\hat{P}\right]_{\times} & \hat{R}^{\top}
	\end{array}\right]\mathbf{U}\nonumber \\
	W= & \left[\begin{array}{cc}
	\hat{R} & 0\\
	\left[\hat{P}\right]_{\times}\hat{R} & \hat{R}
	\end{array}\right]^{\top}\mathbf{U}\label{eq:Gradient}
	\end{align}
	with $\boldsymbol{K}=1+k_{\text{op}}\in\mathbb{R}_{+}$, $k_{\text{op}}$
	being a nonnegative constant to be designed in the following Section,
	$\gamma$ being a positive constant, and $\hat{b}$ being the estimates
	of $b$. 
	\begin{rem}
		\label{rem:Remark2}\cite{hashim2019SE3Det,mohamed2019filters} The
		classic design of nonlinear filters on $SE\left(3\right)$ \cite{hua2011observer}
		select the gain $\boldsymbol{K}$ to be a positive constant. The weakness
		of such an approach is that smaller values of $\boldsymbol{K}$ lead
		to slower transient performance with high measures of robustness in
		the steady-state (less oscillatory performance). In contrast, greater
		values of $\boldsymbol{K}$ results of the faster transient performance
		with less robustness measures in the steady-state (higher oscillation).
	\end{rem}
	Consistent with Remark \ref{rem:Remark2}, the aim is to tune $\boldsymbol{K}$
	to be large at a large error and small at a small error which could
	lead to 1) fast convergence capabilities, and 2) high measures of
	robustness.
	
	\section{Proposed Approach\label{sec:Filter_Strategy}}
	
	In consistence with Remark \ref{rem:Remark2}, $\boldsymbol{K}$ must
	be set large enough at large values of error and small enough at small
	values of error. Hence, fuzzy logic controller (FLC) will be utilized
	to tune $\boldsymbol{K}$ in accordance with the error in pose. The
	basic structure of FLC is composed of 1) fuzzification, which includes
	the input membership function, 2) rule base, and 3) defuzzification,
	which includes the output membership function. Aiming to achieve robust
	and fast adaptation, the parameters of input and output membership
	functions of the FLC will be selected using the gravitational search
	algorithm (GSA) algorithm.
	
	\subsection{Gravitational Search Algorithm\label{subsec:GSA}}
	
\begin{figure}
		\centering{}\includegraphics[scale=0.6]{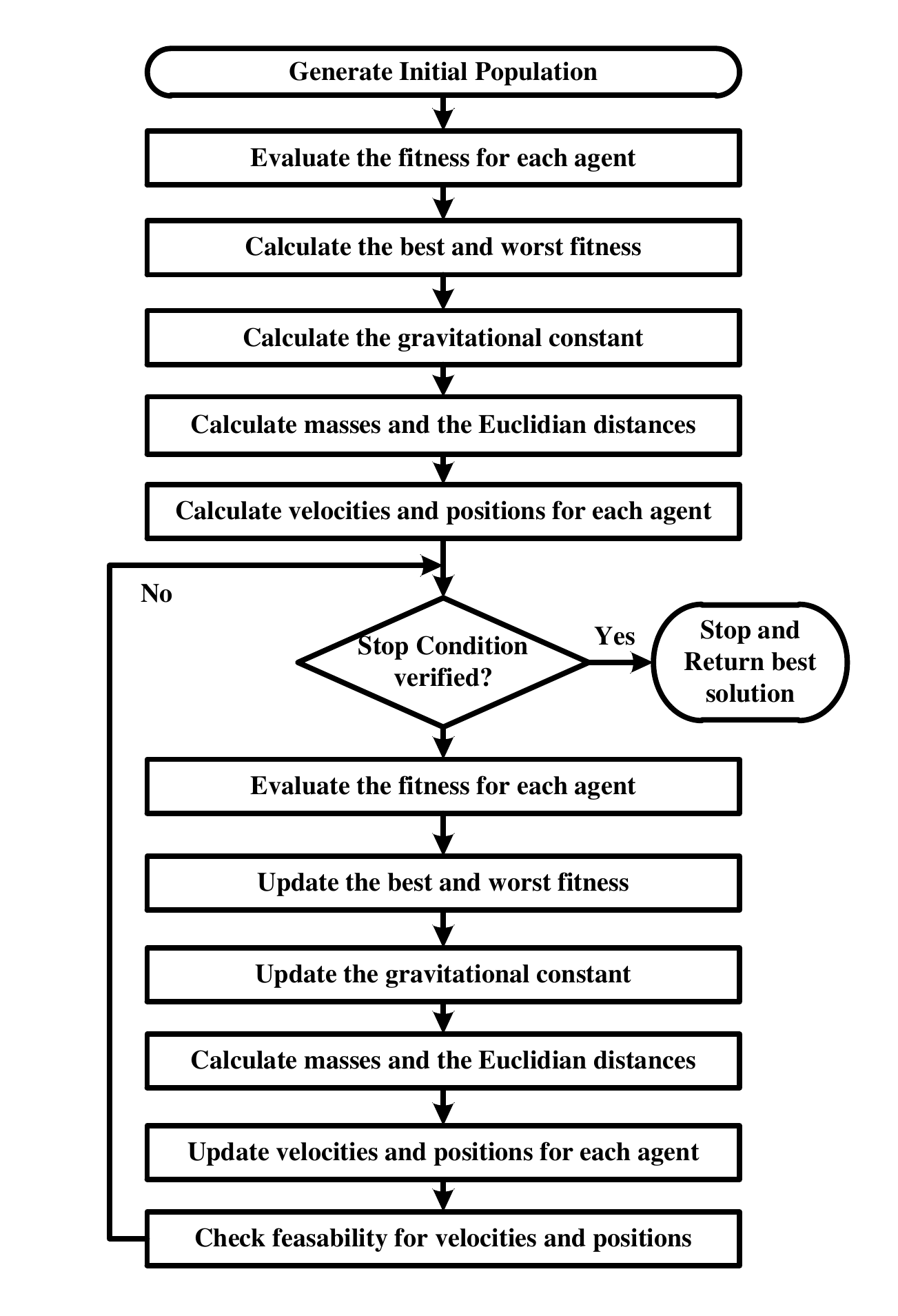}\vspace{-0.3cm}
		\caption{Graphical illustration of GSA algorithm}
		\label{fig:GSA_algorithm}
	\end{figure}

	GSA is an analytical technique, first introduced in 2009 based on
	Newton's laws of universal gravitation \cite{rashedi2009gsa}. The
	algorithm is aligned to the inductive reasoning of gravitational law:
	``for any two objects, every object is attracted to the other object
	by a force which is directly proportional to their mass and inversely
	proportional to their square distance''. Based on gravity principle,
	the gravitational force among any two nodes is 
	
	\begin{equation}
	F\left(t\right)=G\left(t\right)\frac{M_{1}M_{2}}{D\left(t\right)^{2}}\label{eq:GSA_Force}
	\end{equation}
	with $M_{1}$ and $M_{2}$ being masses of node 1 and 2, respectively,
	$D=\left\Vert X_{j},X_{k}\right\Vert ^{2}+\delta$ denoting the Euclidean
	distance between two nodes $i$ and $k$, and $\delta$ denoting a
	small positive constant. $G$ is a gravitational constant and at time
	$t$ is given by
	\begin{equation}
	G\left(t\right)=G\left(t_{0}\right)\exp\left(-\alpha t/T_{\text{final}}\right)\label{eq:GSA_G_Constant}
	\end{equation}
	where $G\left(t_{0}\right)$ denotes the gravitational constant at
	$t_{0}$, $\alpha$ denotes a positive constant, and $T_{\text{final}}$
	denotes final search time. Actually, $T_{\text{final}}$ represents
	the total number of iterations in the search. The gravitational constant
	at time $t$ and the initial  masses have a major role in the cost
	function value. Heavier mass indicates a better node. Similarly, lighter
	mass refers to a worse node. Define a new variable relative to the
	$j$th node
	\begin{equation}
	m_{j}\left(t\right)=\frac{\mathcal{C}_{j}\left(t\right)-\mathcal{C}_{j}^{\text{worse}}\left(t\right)}{\mathcal{C}_{j}^{\text{best}}\left(t\right)-\mathcal{C}_{j}^{\text{worse}}\left(t\right)}\label{eq:GSA_m}
	\end{equation}
	with $N$ being total number of nodes $\forall j=1,2,\cdots,N$, $\mathcal{C}_{j}\left(t\right)$
	being the $j$th cost function at iteration $t$, $\mathcal{C}_{j}^{\text{worse}}\left(t\right)$
	being the worst cost function (highest value), and $\mathcal{C}_{j}^{\text{best}}\left(t\right)$
	being the best cost function (smallest value) in the search process.
	Accordingly, the total mass of the $j$th node is given by
	\begin{equation}
	\overline{M}_{j}\left(t\right)=\frac{m_{j}\left(t\right)}{\sum_{j=1}^{N}m_{j}\left(t\right)}\label{eq:GSA_M}
	\end{equation}
	The acceleration of the node is defined as below
	\begin{equation}
	a_{j,k}\left(t\right)=\frac{F_{j,k}\left(t\right)}{\overline{M}_{j}\left(t\right)}\label{eq:GSA_a}
	\end{equation}
	for all $j=1,2,\cdots,N$ and $k=1,2,\ldots,P$ such that $N$ denotes
	the total number of nodes and $P$ denotes the total number of parameters
	within a single node to be optimized. Also, $F_{j,k}\left(t\right)$
	denotes the force of a particle at position $x_{j,k}\left(t\right)$
	and $\overline{M}_{j}\left(t\right)$ is the mass of the $j$th particle.
	The velocity associated with parameter $k$ in node $j$ at iteration
	$t+1$ is defined by
	\begin{equation}
	\vartheta_{j,k}\left(t+1\right)={\rm rand}_{j}\vartheta_{j,k}\left(t\right)+a_{j,k}\left(t\right)\label{eq:GSA_v}
	\end{equation}
	with ${\rm rand}_{j}$ being a random number between 0 and 1. Finally,
	the position of parameter $k$ in node $j$ at iteration $t+1$ is
	given by
	\begin{equation}
	x_{j,k}\left(t+1\right)=x_{j,k}\left(t\right)+\vartheta_{j,k}\left(t+1\right)\label{eq:GSA_x}
	\end{equation}
	It is worth mentioning that a small set $K_{\text{best}}$ is used
	to contain the best solution over the whole search process. At the
	end of every iteration, the small set $K_{\text{best}}$ is updated.
	The complete flow chart of GSA is illustrated in Fig. \ref{fig:GSA_algorithm}.
	
	\subsection{Pose Filter: GSA and Optimal Fuzzy-tuning}
	
	FLC is a well-known approach that is widely used in various control
	applications. In this work, FLC is employed to control the nonlinear
	pose filter through fine tuning the feedback filter gain. The tuning
	of the filter would contribute to solving the bottleneck between fast
	convergence of pose error and robustness.
	
\begin{figure}
		\vspace{-0.1cm}
		\centering{}\includegraphics[scale=0.6]{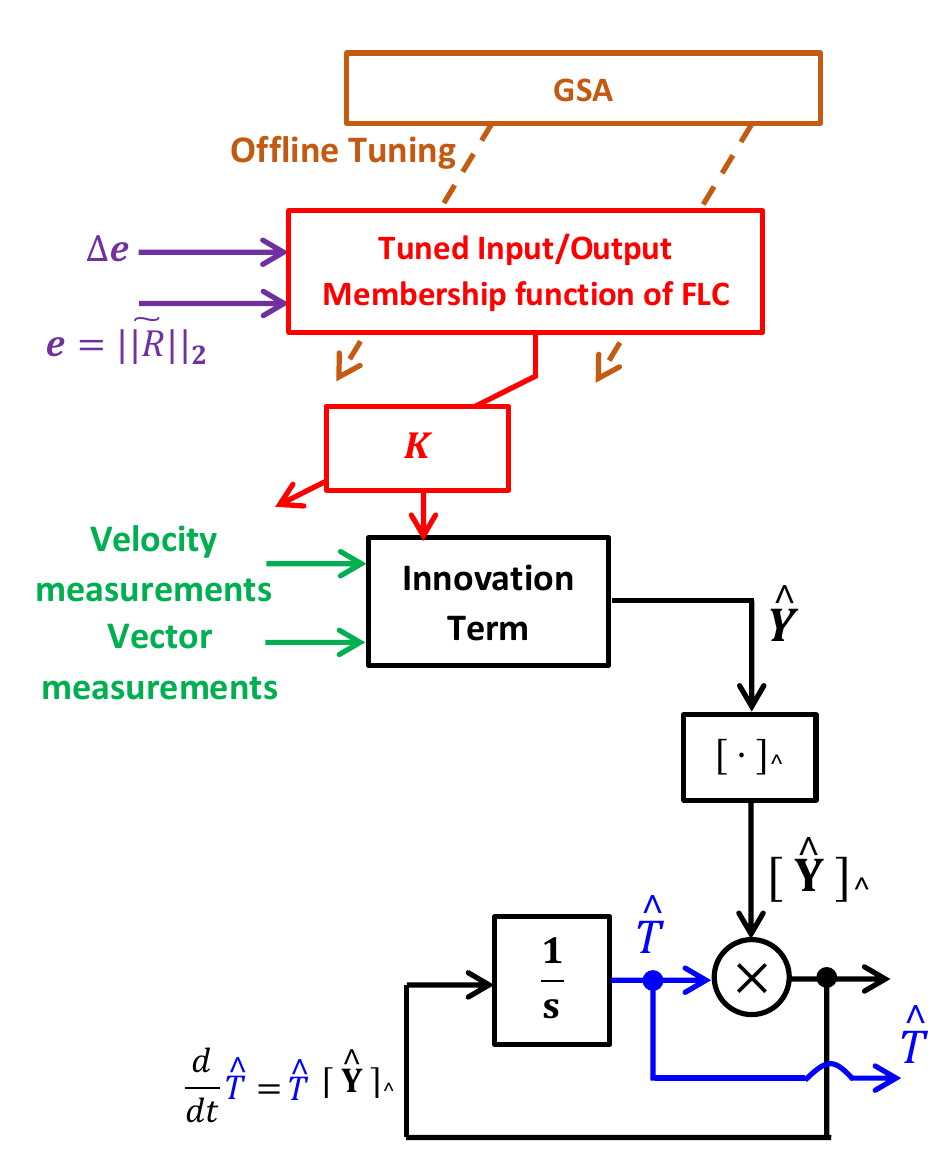}\vspace{-0.2cm}
		\caption{Proposed filter diagram}
		\label{fig:Complete}
	\end{figure}

	The main objective of this work is constructing input and output membership
	functions for FLC that are able to reduce the pose error. This would
	be achieved through the iterative process. The membership functions
	are set to be triangular for both fuzzy inputs and outputs. Each membership
	function has five linguistic variables, namely, very large ($VL$),
	large ($L$), medium ($M$), small ($S$) and very small ($VS$).
	Thus, the setting parameters of the input and output membership function
	are selected through the optimization process. The optimization of
	input and output membership functions values is done using GSA in
	Subsection \ref{subsec:GSA}. See the rule base of the proposed filter
	in Table \ref{tab:FLC}. 
	\begin{table}
		\caption{\label{tab:FLC}Rule base of FLC.}
		
		\centering{}%
		\begin{tabular}{|c||c|c|c|c|c|}
			\hline 
			\noalign{\vskip\doublerulesep}
			$\Delta e\backslash e$ & $VL$ & $L$ & $M$ & $S$ & $VS$\tabularnewline[\doublerulesep]
			\hline 
			\hline 
			\noalign{\vskip\doublerulesep}
			$VL$ & $VL$ & $VL$ & $VL$ & $V$ & $V$\tabularnewline[\doublerulesep]
			\hline 
			\noalign{\vskip\doublerulesep}
			$V$ & $VL$ & $VL$ & $VL$ & $V$ & $M$\tabularnewline[\doublerulesep]
			\hline 
			\noalign{\vskip\doublerulesep}
			$M$ & $VL$ & $VL$ & $V$ & $M$ & $M$\tabularnewline[\doublerulesep]
			\hline 
			\noalign{\vskip\doublerulesep}
			$S$ & $VL$ & $VL$ & $M$ & $M$ & $S$\tabularnewline[\doublerulesep]
			\hline 
			\noalign{\vskip\doublerulesep}
			$VS$ & $VL$ & $VL$ & $M$ & $S$ & $VS$\tabularnewline[\doublerulesep]
			\hline 
		\end{tabular}
	\end{table}
	The $j$th cost function is selected as below
	\begin{align}
	\mathcal{C}_{j} & =0.3\times e_{tr}+e_{ss}\nonumber \\
	e_{tr} & =\sum_{0\leq t\leq1}||\tilde{R}||_{I}+0.2\times\sum_{0\leq t\leq1}||\tilde{P}\left(t\right)||\nonumber \\
	e_{ss} & =\sum_{4\leq t\leq14}||\tilde{R}||_{I}+0.2\times\sum_{4\leq t\leq14}||\tilde{P}\left(t\right)||\label{eq:Cost_function}
	\end{align}
	with $e_{tr}$ being the transient time over the period of $0$ to
	$1$ seconds, $e_{ss}$ being the steady-state error over the period
	of $4$ to $15$ seconds for a sampling time of $0.01$ seconds. $0.3$
	and $0.2$ in Eq. \eqref{eq:Cost_function} are weighting factors
	and are selected after a set of trials. The input and output membership
	functions have constraint values given in Eq. \eqref{eq:Inp_MMF_Constraints}
	and Eq. \eqref{eq:Out_MMF_Constraints}, respectively
	
	\begin{equation}
	\begin{cases}
	\left[0,0,0\right] & \leq\left[0,0,k_{1}\right]\leq\left[0,0,0.15\right]\\
	\left[0,0,0.1\right] & \leq\left[k_{2},k_{3},k_{4}\right]\leq\left[0.2,0.2,0.2\right]\\
	\left[0.05,0.1,0.1\right] & \leq\left[k_{5},k_{6},k_{7}\right]\leq\left[0.2,0.3,0.4\right]\\
	\left[0.1,0.2,0.3\right] & \leq\left[k_{8},k_{9},k_{10}\right]\leq\left[0.4,0.8,0.8\right]\\
	\left[0.2,1,1\right] & \leq\left[k_{11},1,1\right]\leq\left[0.7,1,1\right]
	\end{cases}\label{eq:Inp_MMF_Constraints}
	\end{equation}
	\begin{equation}
	\begin{cases}
	\left[0,0,0\right] & \leq\left[0,0,k_{12}\right]\leq\left[0,0,10\right]\\
	\left[0,0,5\right] & \leq\left[k_{13},k_{14},k_{15}\right]\leq\left[10,20,30\right]\\
	\left[5,10,20\right] & \leq\left[k_{16},k_{17},k_{18}\right]\leq\left[20,50,50\right]\\
	\left[20,20,40\right] & \leq\left[k_{19},k_{20},k_{21}\right]\leq\left[50,70,90\right]\\
	\left[30,100,100\right] & \leq\left[k_{22},100,100\right]\leq\left[70,100,100\right]
	\end{cases}\label{eq:Out_MMF_Constraints}
	\end{equation}
	where $k_{1}$ to $k_{22}$ are parameters of the membership functions
	to be optimized using GSA with respect to the cost function defined
	in Eq. \eqref{eq:Cost_function} in addition to the constraints in
	Eq. \eqref{eq:Inp_MMF_Constraints} and Eq. \eqref{eq:Out_MMF_Constraints}.
	Fig. \ref{fig:Complete} illustrates the complete diagram of the proposed
	filter strategy.
	
	\section{Simulation\label{sec:Results}}
	
	\subsection{Pose Measurements and Initialization}
	
	Consider the following set of measurements 
	\[
	\begin{cases}
	\Omega_{m} & =\Omega+b_{\Omega}+n_{\Omega}\text{ (rad/sec)}\\
	V_{m} & =V+b_{V}+n_{V}\text{ (m/sec)}\\
	\Omega & =\left[\sin\left(0.7t\right),0.7\sin\left(0.5t+\pi\right),0.5\sin\left(0.3t+\frac{\pi}{3}\right)\right]^{\top}\\
	V & =0.3\left[\sin\left(0.6t\right),\sin\left(0.4t\right),\sin\left(0.1t\right)\right]^{\top}\\
	b_{\Omega} & =0.1\left[1,-1,1\right]^{\top},\hspace{1em}\hspace{1em}n_{\Omega}=\mathcal{N}\left(0,0.2\right)\\
	b_{V} & =0.1\left[2,5,1\right]^{\top},\hspace{1em}\hspace{1em}n_{V}=\mathcal{N}\left(0,0.1\right)\\
	{\rm v}_{i}^{\mathcal{B}} & =R^{\top}{\rm v}_{i}^{\mathcal{I}}+b_{i}^{\mathcal{B}}+n_{i}^{\mathcal{B}},\hspace{1em}{\rm v}_{3}^{\mathcal{B}}={\rm v}_{1}^{\mathcal{B}}\times{\rm v}_{2}^{\mathcal{B}}\\
	{\rm v}_{1}^{\mathcal{I}} & =\frac{1}{\sqrt{3}}\left[1,-1,1\right]^{\top},\hspace{1em}{\rm v}_{2}^{\mathcal{I}}=\left[0,0,1\right]^{\top}\\
	b_{1}^{\mathcal{B}} & =0.1\left[1,-1,1\right]^{\top},\hspace{1em}\hspace{1em}n_{1}^{\mathcal{B}}=\mathcal{N}\left(0,0.1\right)\\
	b_{2}^{\mathcal{B}} & =0.1\left[0,0,1\right]^{\top},\hspace{1em}\hspace{1em}n_{2}^{\mathcal{B}}=\mathcal{N}\left(0,0.1\right)\\
	y_{1}^{\mathcal{B}} & =R^{\top}\left(p_{1}^{\mathcal{I}}-P\right)+\bar{b}_{1}^{\mathcal{B}}+\bar{n}_{1}^{\mathcal{B}}\\
	p_{1}^{\mathcal{I}} & =\left[0.5,\sqrt{2},1\right]^{\top},\hspace{1em}\bar{n}_{1}^{\mathcal{B}}=\mathcal{N}\left(0,0.1\right)
	\end{cases}
	\]
	where $n=\mathcal{N}\left(0,0.1\right)$ is a random noise vector
	with $0$ mean and standard deviation of $0.1$. Also, for very large
	error, consider the following:
	\[
	\boldsymbol{T}\left(0\right)=\mathbf{I}_{4},\hspace{1em}\hat{\boldsymbol{T}}\left(0\right)=\left[\begin{array}{cccc}
	-0.829 & 0.293 & 0.343 & 4\\
	0.399 & 0.157 & 0.903 & -3\\
	0.210 & 0.943 & -0.257 & 5\\
	0 & 0 & 0 & 1
	\end{array}\right]
	\]

	\subsection{GSA Implementation}
	
\begin{figure}
		\centering{}\includegraphics[scale=0.21]{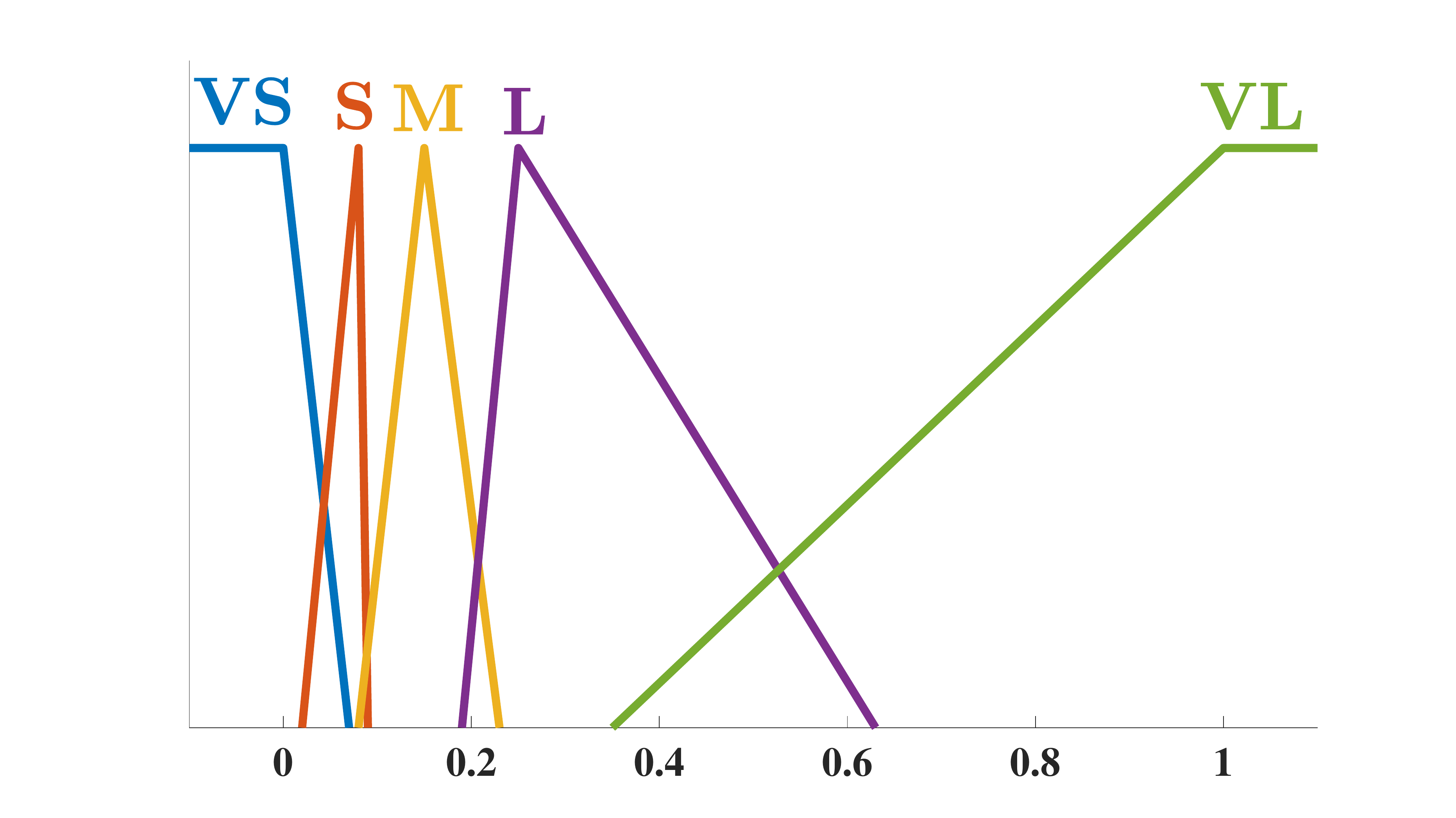}\vspace{-0.3cm}
		\caption{Error and rate of error membership functions}
		\label{fig:FLC_Inp_MMF}
	\end{figure}

\begin{figure}
		\centering{}\includegraphics[scale=0.21]{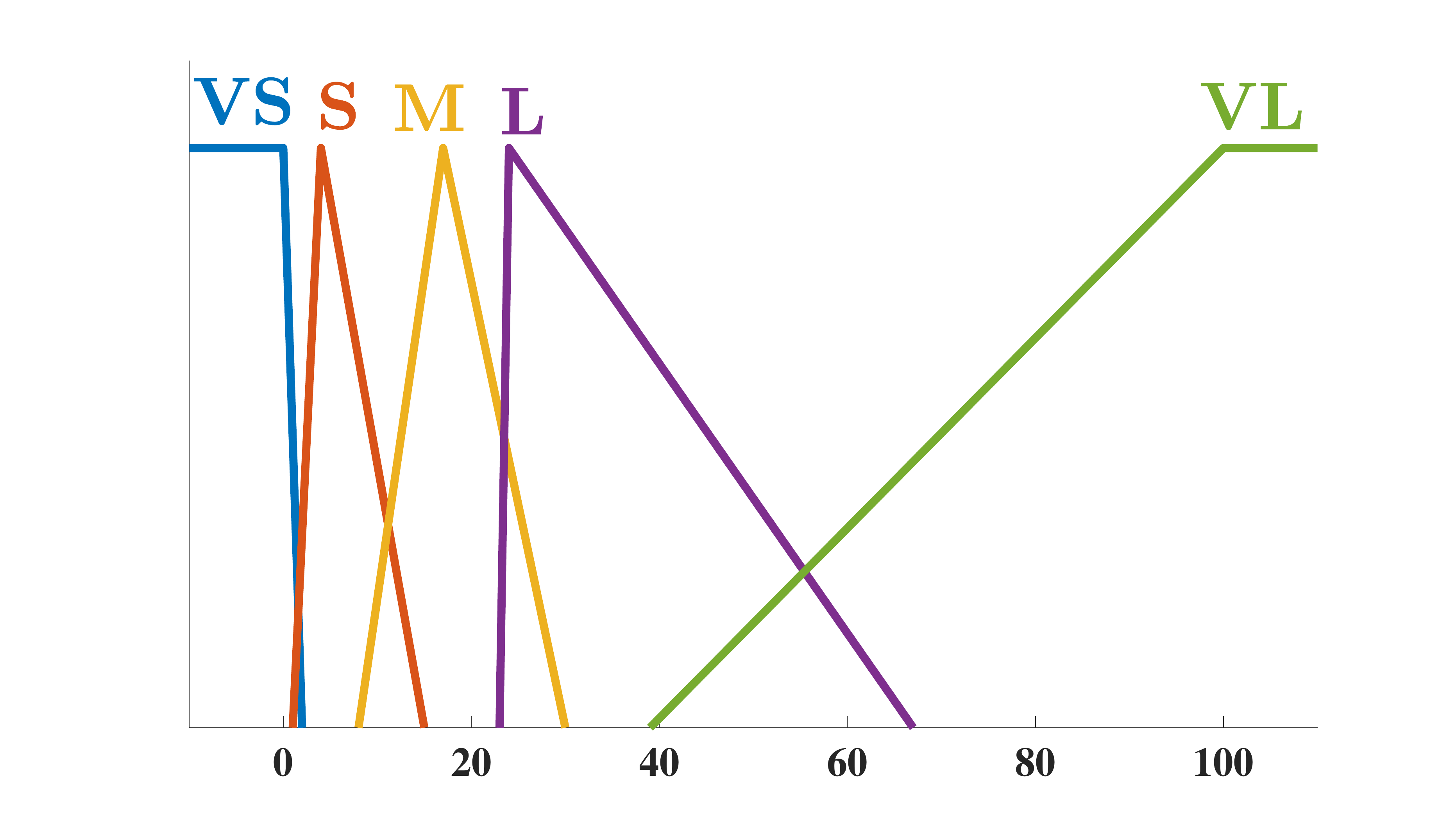}\vspace{-0.3cm}
		\caption{Output membership function}
		\label{fig:FLC_Out_MMF}
	\end{figure}

	For the implementation, Eq. \eqref{eq:GSA_x} represents the position
	of the particle and Eq. \eqref{eq:GSA_m} denotes mass with respect
	to the quality of the cost function. $N$ is the number of nodes in
	the space. The total number of iterations is 250. The number of nodes
	to be allocated is $N=100$. Within every node, there are $22$ parameters
	to be optimized, $k_{1}$ to $k_{22}$, given in Eq. \eqref{eq:Inp_MMF_Constraints}
	and \eqref{eq:Out_MMF_Constraints}. Fig. \ref{fig:FLC_Inp_MMF} and
	\ref{fig:FLC_Out_MMF} represent the optimized input and output membership
	function after completing the search process.
	
	\vspace{-0.3cm}
	
	\subsection{Robustness of the Proposed Approach}
	
\begin{figure}
		\centering{}\includegraphics[scale=0.32]{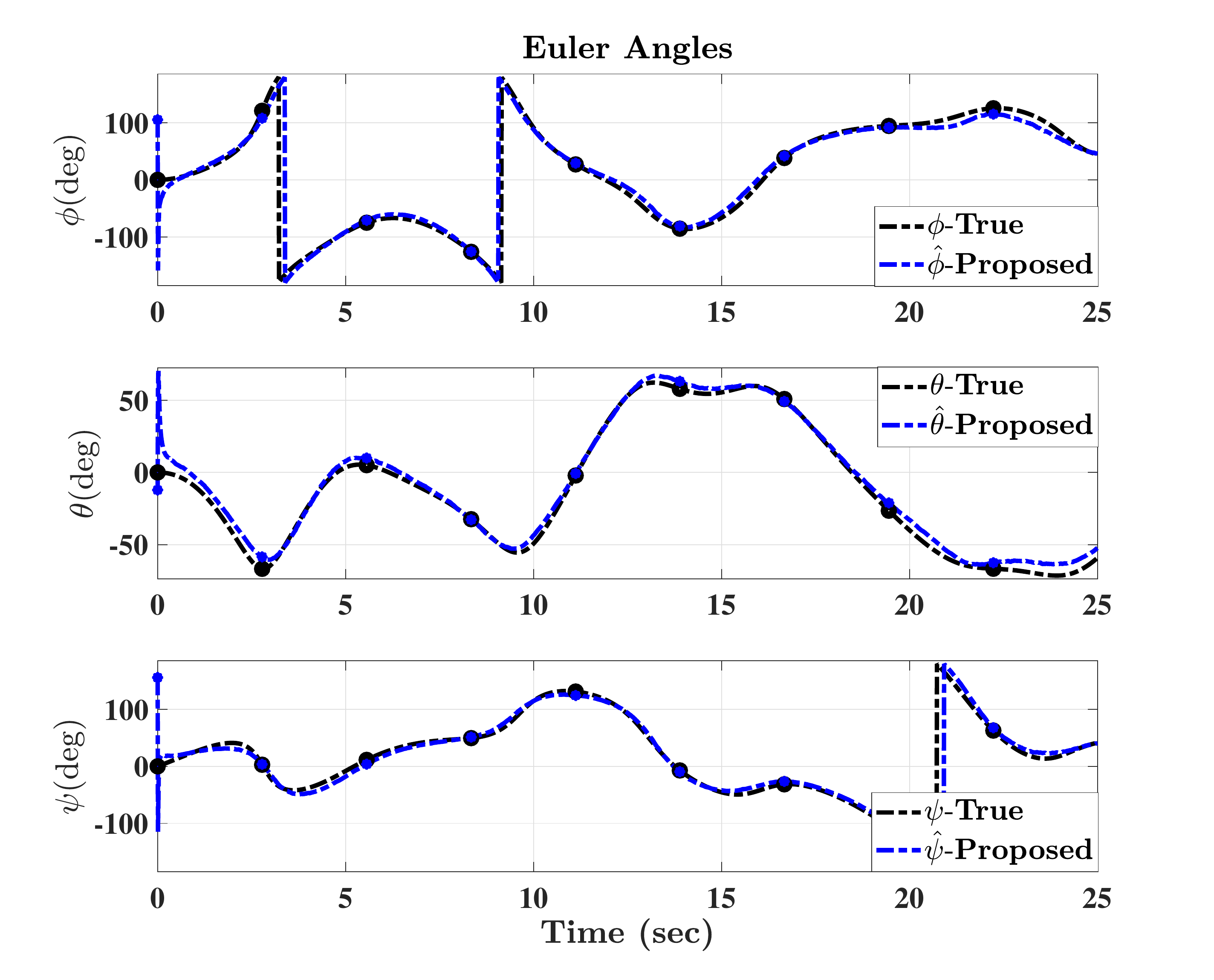}\vspace{-0.3cm}
		\caption{Euler angles ($\phi,\theta,\psi$): True vs Estimate (Proposed)}
		\label{fig:Euler}
	\end{figure}

\begin{figure}
		\vspace{-0.3cm}
		\centering{}\includegraphics[scale=0.26]{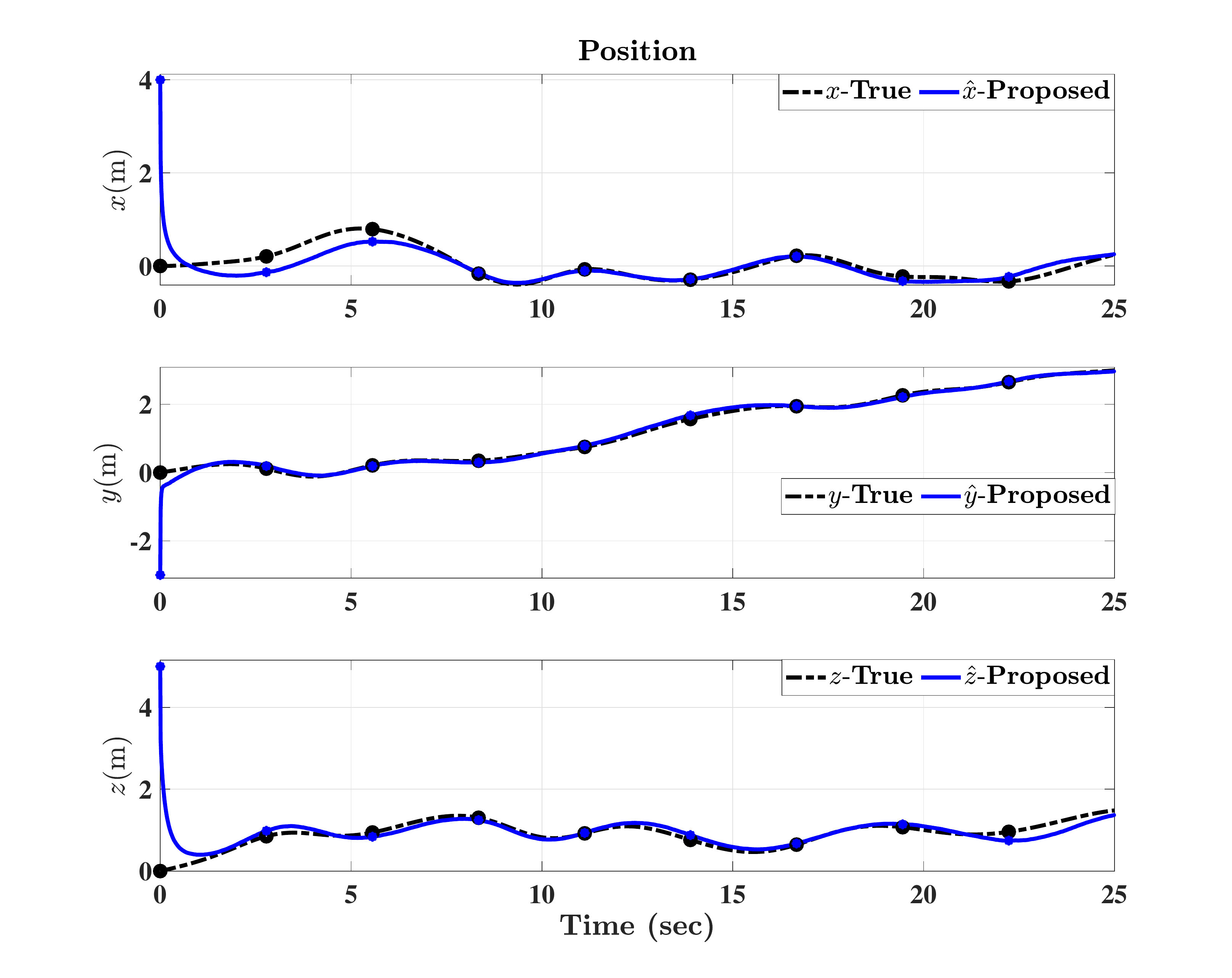}\vspace{-0.3cm}
		\caption{Position ($x,y,z$): True vs Estimate (Proposed)}
		\label{fig:P}
	\end{figure}

\begin{figure}
		\vspace{-0.3cm}
		\centering{}\includegraphics[scale=0.23]{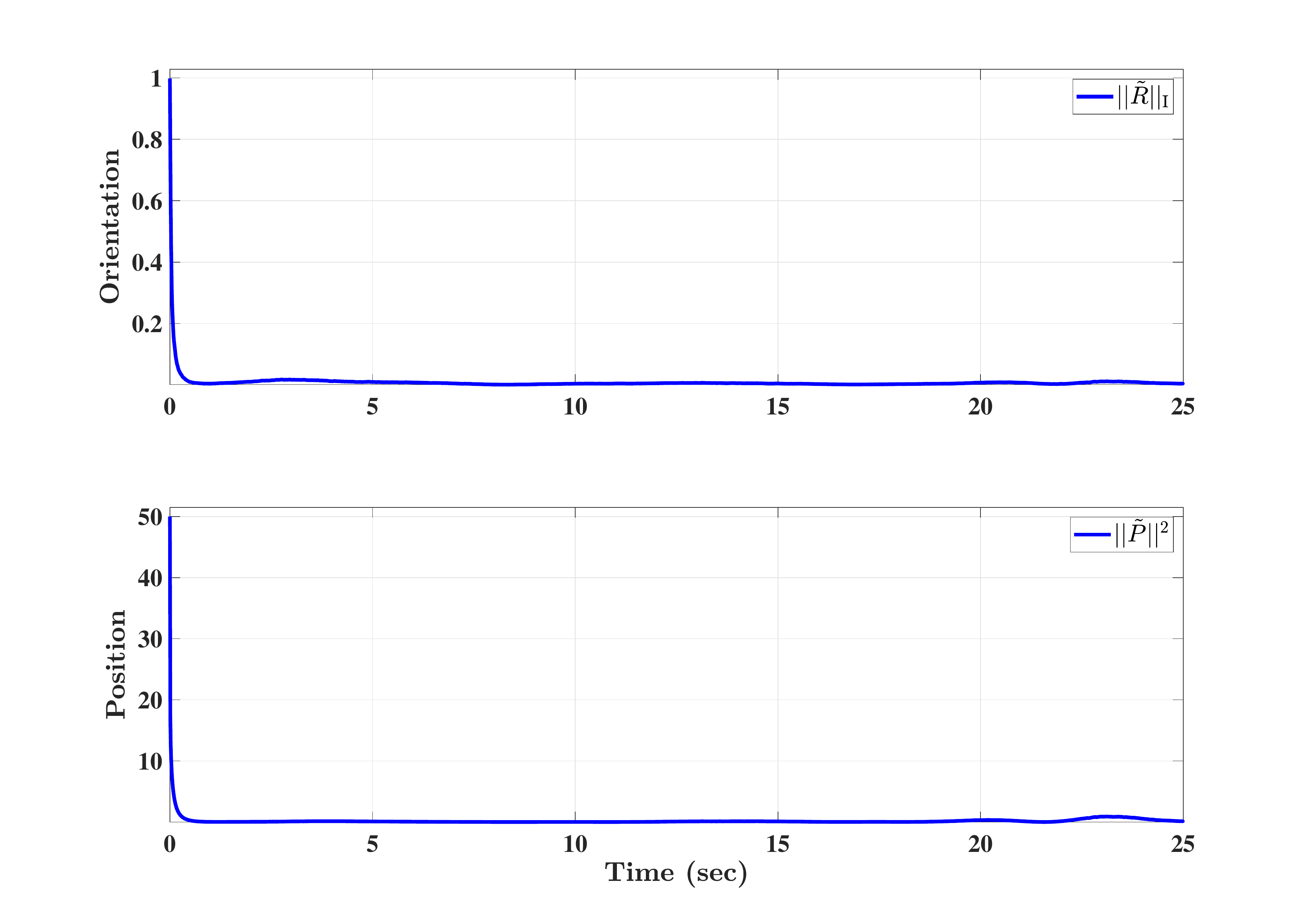}\vspace{-0.3cm}
		\caption{Normalized Error: Position \& Euclidean distance}
		\label{fig:Error}
	\end{figure}

	Fig. \ref{fig:Euler} and \ref{fig:P} reveal smooth and fast tracking
	convergence of estimated Euler angles versus the true Euler angles
	$\left(\phi,\theta,\psi\right)$ and estimated position versus the
	true position, respectively. The proposed approach shows impressive
	tracking performance against uncertain measurements and large error
	in initialization. This can be confirmed by the results presented
	in Fig. \ref{fig:Error} that show how the filter initiated at large
	values of error and converged very close to the origin in a short
	time. As such, Fig. \ref{fig:Error} reveals that the proposed approach
	is characterized with fast adaptation and robustness.
	

	\section{Conclusion\label{sec:SO3PPF_Conclusion}}
	
	This paper presents the fuzzy logic controller (FLC) design, tuned
	with the gravitational search algorithm (GSA) for a nonlinear pose
	filter. GSA optimization has been utilized to find the optimal parameters
	of the input and output membership functions of the FLC. The proposed
	approach tunes the adaptation gain on-line, to allow for fast adaptation.
	In addition, owing to the smooth tuning, the proposed filter maintains
	a high measure of robustness. Simulation results illustrate fast convergence
	of the pose (attitude and position) error and robustness against large
	initialization error and uncertain measurements. In the future, there
	is a plan to implement the proposed approach on a real module and
	compare it against existing algorithms in the literature. Also, we
	are planning to compare GSA against other methods of evolutionary
	techniques.
	
	\section*{Acknowledgment}
	
	The authors would like to thank \textbf{Maria Shaposhnikova} for proofreading
	the article.
	
	%\ifCLASSOPTIONcaptionsoff %\bibliographystyle{spmpsci}      % mathematics and physical sciences
	\bibliographystyle{IEEEtran}
	\bibliography{arXiv_Pose_Conf}
	% name your BibTeX data base

\end{document}